\documentclass[12pt]{revtex4}
\usepackage{ulem}
\usepackage{url}
\usepackage{epsfig}
\usepackage{graphicx,color}
\usepackage[psamsfonts]{amssymb}
\usepackage{amsmath}
\usepackage{indentfirst}

\usepackage{etoolbox}

\usepackage[latin1,applemac]{inputenc}

\newcommand{\be}{\begin{equation}}
\newcommand{\ee}{\end{equation}}
\newcommand{\ba}{\begin{eqnarray}}
\newcommand{\ea}{\end{eqnarray}}

\begin{document}
\title{How Much is the Whole Really More than the Sum of its Parts? $1 \boxplus 1 = 2.5$: Superlinear Productivity in Collective Group Actions}

\author{D. Sornette}
\email{dsornette@ethz.ch}
\affiliation{Department of Management, Technology and Economics,
ETH Zurich, Scheuchzerstrasse 7, CH-8092 Zurich, Switzerland}

\author{T. Maillart}
\email{thomas.maillart@ischool.berkeley.edu}
\affiliation{School of Information, \\UC Berkeley, 102 S Hall Rd, Berkeley, CA 94704, United States}

\author{G. Ghezzi}
\email{ghezzi@ifi.uzh.ch}
\affiliation{Department of Informatics, University of Zurich, Binzmühlestrasse 14, CH-8050 Zurich, Switzerland}

\date{\today}

\begin{abstract}
\vskip 1cm
In a variety of open source software projects, we document a superlinear growth of production ($R \sim c^\beta$) as a function of the number of active developers $c$, with $\beta \simeq 4/3$ with large dispersions. For a typical project in this class, doubling of the group size multiplies typically the output by a factor $2^\beta=2.5$, explaining the title. This superlinear law is found to hold for group sizes ranging from 5 to a few hundred developers. We propose two classes of mechanisms, {\it interaction-based} and {\it large deviation}, along with a cascade model of productive activity,
which unifies them. In this common framework, superlinear productivity requires
that the involved social groups function at or close to criticality, in the sense of a subtle balance
between order and disorder. We report the first empirical test of the renormalization of the exponent
of the distribution of the sizes of  first generation events into the renormalized exponent of the distribution of clusters resulting from the cascade of triggering 
over all generation in a critical branching process
in the non-meanfield regime. Finally, we document a size effect in the strength and variability of the superlinear effect, with smaller groups exhibiting widely distributed superlinear exponents, some of
them characterizing highly productive teams. In contrast, large groups tend to have a smaller superlinearity and less variability.
\end{abstract}
\maketitle

\pagebreak

\section{Introduction}
Since at least Aristotle, the adage in the title has permeated human thinking, with prominent influence 
in psychology (Gestalt theory \cite{humphrey1924psychology}), biology (brain functions \cite{damoiseaux2009}, 
ecological networks \cite{jorgensen2012}), physics 
(spontaneous symmetry breaking \cite{anderson1963} and the ``more is different'' concept \cite{anderson1972}),
economics \cite{arthur1994increasing,krugman1996} among a wealth of other examples.  
Prominent among other developments are
the fields of complexity science, synergetics and complex adaptive system theory, which
strive to understand natural and social systems in terms of a systemic or holistic approach,
where the above adage is translated into the scientific concept of {\it emergence} that  
results from repetitive interactions between simple
constituting elements in extended out-of-equilibrium adaptive systems.  
Dealing with groups such as firms and production units, 
management science also strives to understand when and how a group 
can be more than the sum of individuals, and to design ways to improve team performance
\cite{tziner1985,sundstrom1990work,cohen1997,neuman1999team}, 
through the mechanism of complementarity in organization
\cite{ennen2010,lin2006communities} and innovations \cite{sacramento2006team}. 
Because most activities in our modern environment require coordination and
collaborative actions within groups of widely varying sizes, 
it is the dream of any manager, be it in the public or private sector, to find the 
gears that could enhance productivity. 

Notwithstanding their importance in human culture
and civilization since ancient times, we still have a limited
understanding of the mechanisms at the origin of group productivity.
Moreover, we do not really understand the conditions under which 
the whole is more than the sum of its parts, and how to quantify its productivity with 
respect to its different constituents.
The bottlenecks hindering progress include the difficulties for quantifying productivity
as well as the obstacles of controlled experiments that allow for clean conclusions.
Indeed, most human groups and systems are entangled in their functioning and
objectives, and are rarely amenable to systematic and continuous observations
suitable for rigorous scientific analyses.

To address these problems, we use a source of data in which group cooperation 
is ubiquitous and can be quantified in great details, namely 
the dynamics of production during the development of open source software (OSS) projects.
Because OSS development is essentially collective, iterative, and cumulative, 
and the overhead costs for interactions is small thanks to the cheap electronic
support mediating exchanges between developers, the study of potential 
increases of productivity by interaction and cooperation between
several contributing developers is particularly well suited.

Section \ref{sdjkfsohtoe} presents the main empirical evidence of the superlinear production law
found for open source software projects.  Section \ref{sfghdugeuijvnei} presents two classes 
of mechanisms at the origin of superlinear production, which are 
unified in the cascade model of productive activity. Empirical data tests are found to
support the model. Section \ref{sdfuishfqsatug} compares and attempts to reconcile present findings
for OSS and the superlinear law previously reported for cities. Section \ref{sdfudprtptm} discusses our results, and Section \ref{sdfiuetgnzgposfpkn} concludes.

\section{Quantification of productivity in open source software projects}
\label{sdjkfsohtoe}
\begin{figure}[h]
\centerline{\epsfig{figure=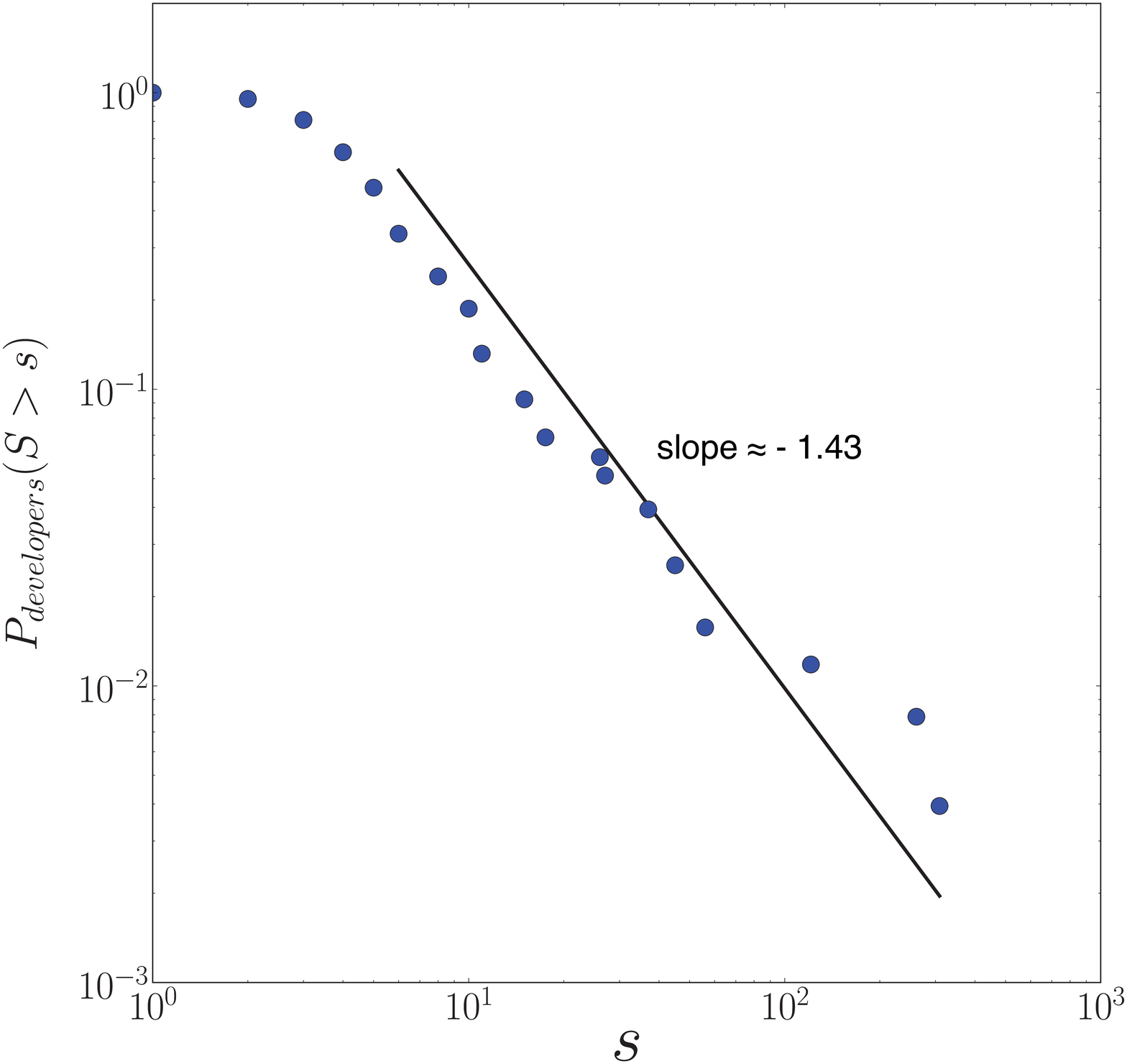,angle=0,width=10cm,scale=1}}
\caption{\footnotesize {Distribution of project sizes in our sample quantified by their total number of developers. The distribution follows approximately a power law with exponent $\alpha \approx 1.4$, 
with an apparent deviation in the tail possibly
resulting from an over-sampling bias of large projects. The bend down for small projects is likely the result
of an under-sampling bias.}}
\label{fig:distribution_project_size}
\end{figure}

We have analyzed the production for $164$ open source software projects of size 
ranging from $5$ to $1678$ contributors.
Figure \ref{fig:distribution_project_size} shows the complementary cumulative distribution of project sizes in our sample
quantified by the number of developers involved in each project. The distribution is an approximate
power law $P_>(S) := {\rm Pr(size}>S) \sim  1/S^\alpha$  with exponent $\alpha \approx 1.4$,
 which reflects a large heterogeneity of project sizes with 
few projects attracting many developers and a multitude of projects with just a few developers. 
The simplest generic mechanism for such power law distribution of human group sizes is
proportional growth coupled with birth and death \cite{saichev2009,malevergne2013zipf}
as verified empirically in OSS package reuse \cite{Maillart2008}, in group \cite{zhang2011} and in product \cite{saichev2013hierarchy} dynamics.

\begin{figure}[h]
\centerline{\epsfig{figure=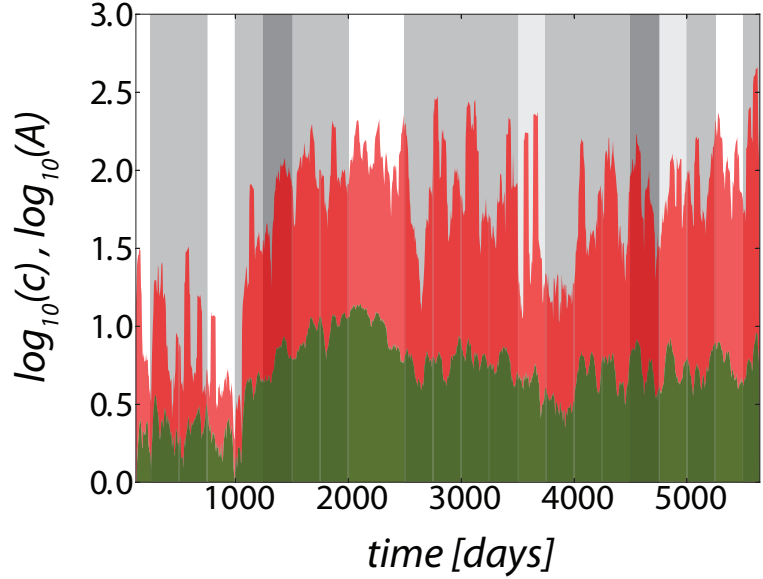,angle=0,width=10cm,scale=1}}
\caption{\footnotesize {Typical time series of open source software development (e.g. Apache Web Server) with active contributors (green area) and their productive activity (red area). For clarity, the time series are represented in logarithmic scale and they have been smoothed with a rolling window of $45$ days. Over the whole project history, various epochs of productive activity can be found. The background grey areas indicate three levels of the productivity exponent $\beta$
defined by equation (\ref{superlineqex}) (light grey for $\beta < 1$, grey for $1 \leqslant \beta < 2$ and dark grey for $\beta \geqslant 2$) for time windows of 250 days. Blank areas show time windows for which $\beta$ could not be fitted, mainly because the numbers of active contributors (resp. commits) were strongly varying over these periods. In other words, it is possible
that super linear production was occurring in these periods but we could not determined it.}}
\label{tseries}
\end{figure}

A first idea would be to quantify the total production 
(for instance proxied by the number of lines of code, commits or the number of packages)
of each software and search for a relationship with the total number of involved developers
over the whole project. This is misleading because the total output results from a complex
interplay between a time varying numbers of involved developers and the intermittent duration
and intensity of their contributions.
In the extreme limit, a single developer working over a lifetime may produce as much as
tens or even hundreds of developers over a few months. The large variability 
of developer numbers and contributions as a function of time for each 
project is illustrated by Figure \ref{tseries}, which shows the intermittent 
dynamics of active contributors as well as their productive activity as a function of time (in logarithmic scales).
 
To capture more faithfully the actions of contributions via cooperation, we
propose to focus on short-term production and group sizes. For each project,
we partition its lifetime in time windows of a fixed size that we shift over the whole project duration.
We then quantify the production in each window and study its relation to the number
of active developers during that same time window. As proxies for the production of developers,
we could use either use lines of codes (${\rm LOCs}$) or commits. ${\rm LOCs}$ are straightforward
metrics but suffer from the criticism that real production and quality is not in general proportional
to the number of code lines. Indeed, excellent contributions are in general characterized by efficient
and elegant coding associated with conciseness. Among software developers, it is well recognized
that the number of LOCs contributed is not a predictor of quality. However, in open collaboration, each innovation step can be seen as a commit uploaded and compounded on an online repository, which keeps track of all changes over time. Each commit reflects the contributor's {\it commitment} to expose to the community her proposed solution to an open problem. Commits are the elementary units 
that get peer-reviewed, tested and eventually integrated in the project knowledge base. 
Thus, they are a direct measure of the iterative productive process at work in peer-production. 
All commit activities are parsimoniously indexed and timestamped on the project repository.

\begin{figure}[h]
\centerline{\epsfig{figure=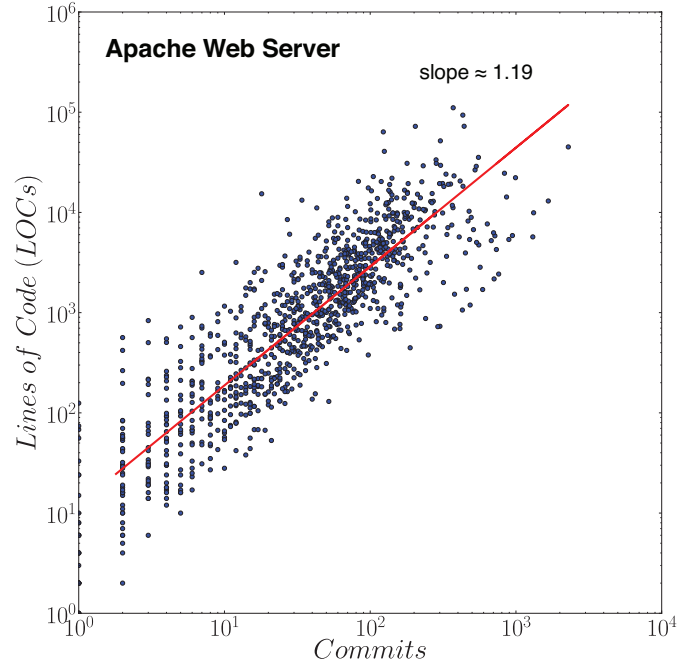,angle=0,width=10cm,scale=1}}
\caption{\footnotesize  Scaling relation ${\rm LOCs} \sim ({\rm Commits})^\delta$ between commits and lines of code. For the Apache Web Server project, the scaling exponent is $\delta = 1.2 \pm 0.2$ ($p < 0.01$, $R^2= 0.87$). 
For the vast majority of projects, the relation between lines of code and commits exhibits the same scaling with $\delta \gtrapprox  1$, suggesting that we can use either commits or lines of codes, as both provide a consistent and therefore
robust measure of contribution (and in addition that commits
may themselves result from cascades of code production.}
\label{fig:locs_vs_commits}
\end{figure}

Notwithstanding these arguments in favor of using commits as metrics of production,
it is useful to test for a possible relation between 
${\rm LOCs}$ and ${\rm Commits}$. Figure \ref{fig:locs_vs_commits}
documents a robust scaling relationship ${\rm LOCs} \sim ({\rm Commits})^\delta$, with 
exponents $\delta \gtrapprox  1$ for most of the projects. 
These findings shown in Figure \ref{fig:locs_vs_commits} bolster our confidence
in the robustness of the findings reported below, which should not be sensitive to the 
specific choice of the metric for production. 

\begin{figure}[h]
\centerline{\epsfig{figure=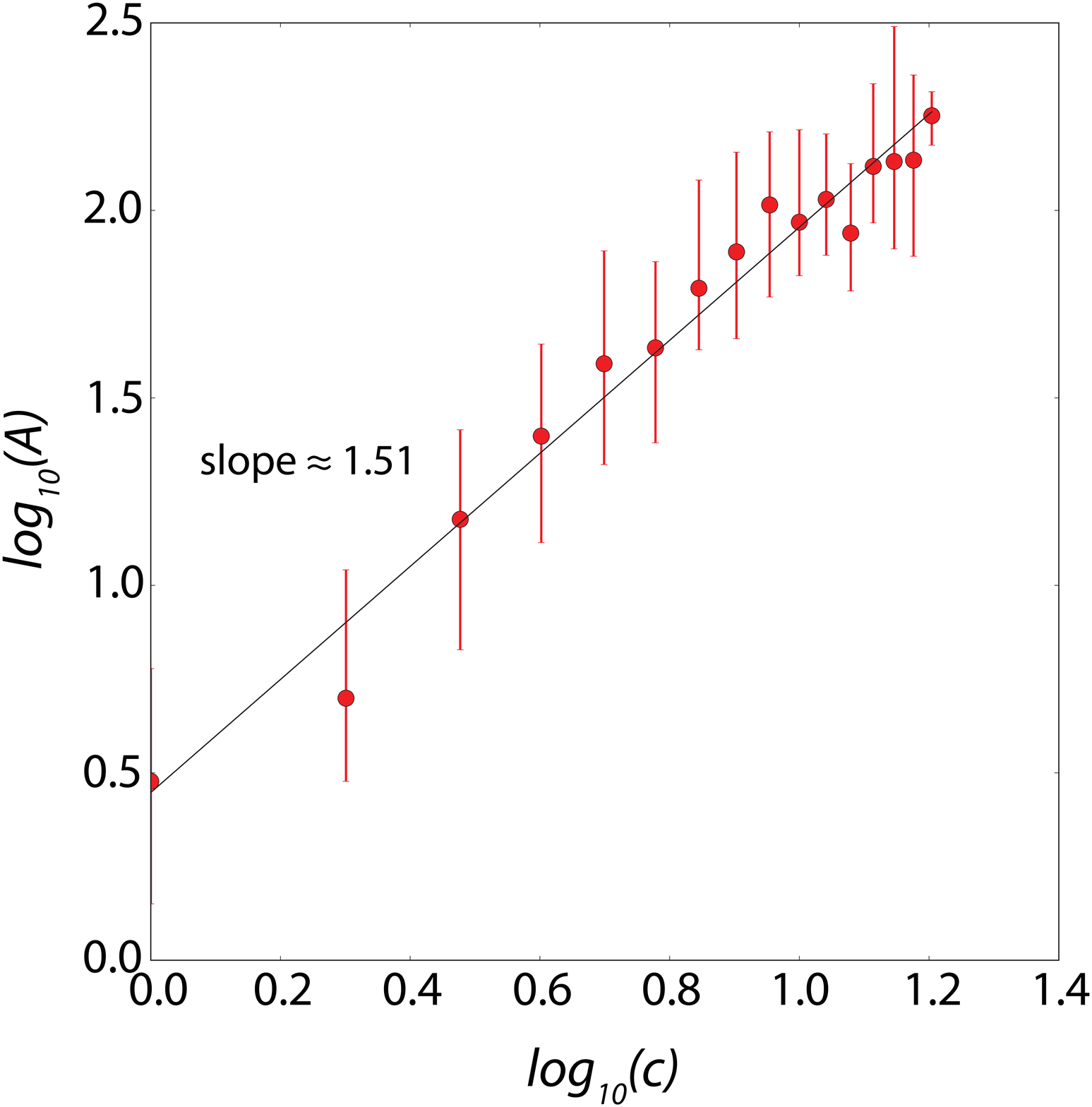,angle=0,width=10cm,scale=1}}
\caption{\footnotesize {Typical superlinear relation in double logarithmic scale of the productive contribution $R$ as a function of active contributors $c$ per 5-day time windows for Apache Web Server ({\it http://httpd.apache.org/}). The scaling exponent $\beta \approx 1.5$ ($p<0.001$ and $R^2=0.99$) is shown as the slope of a straight line in double logarithmic scale. The error bars show the 25th and 75th percentiles of contributors log-bins. Since $2^{1.5} =2.8$, this explains the title of this paper.}}
\label{scaling}
\end{figure}

Figure \ref{scaling} demonstrates the typical superlinear relationship
\begin{equation}
R \sim c^\beta
\label{superlineqex}
\end{equation}
where the production $R$ is defined as the total number of commits measured 
per 5-day time windows for the Apache Web Server (\url{http://httpd.apache.org/})
and $c$ is the number of active contributors in the same 5-day time windows.
Contrary to the naive expectation that the production $R$ should be proportional to the number $c$
of developers, Figure \ref{scaling} documents a superlinear relationship with exponent 
$\beta \approx 1.5 \pm 0.1$, therefore significantly larger than the value $1$ describing a simple
proportionality $R \propto c$. Over all OSS projects studied, 
the estimated statistical average is $\hat\beta \approx 4/3$. Since $2^{4/3} =2.5$, this explains the title of this paper. For many projects, $\beta$ is larger than $4/3$, such as the 
Apache Web Server project shown in figure \ref{scaling}, for which $2^{1.5} =2.8$.

\section{Mechanisms for superlinear production}
\label{sfghdugeuijvnei}

We consider two classes of mechanisms for superlinear production.

\subsection{Interaction-based mechanism for superlinear production}
There is a variety of channels by which
contributors commit more solutions to problems when the community is more active.
The peer-review process is more likely to occur when more contributors are active,  there are incentives to share early with the community to avoid redundant work and some problems require collective intelligence to increase their chance to be solved \cite{woolley2010}, because they require tight coordination among different technical parts of the code \cite{O'Mahony2007}. When interactions occur, the observed increasing return of productive activity implies that the 
change $dR/dc$ of productivity upon the addition of a developer due to
the existence of interactions is not a constant but grows itself with the number of active contributors (as $\sim c^{\beta -1}$ with $\beta>1$). There is thus a remarkable increase of {\it productive activity}, not only as the sum of increased individual commits, but also as a result of interactions among active contributors. In standard models of interaction, linearity is the rule ($\beta =1$) and superlinearity is only obtained when triggering of action is close to {\it criticality} \cite{grimmett1994probability,liggett2005interacting,galam2012sociophysics,domb2000phase}, when one action triggers on average one follow-up action, ensuring that the dynamics remains 
delicately poised between growth and decay. Criticality is indeed characterized by a superlinear
dependence of the response function as a function of the driving field. Interpreted in the present context,
the response function is the total production and the driving field is the number of active developers.
Therefore, an explanation of superlinear productivity by the interaction-based mechanism requires 
elucidating under which circumstances open source projects operate close to or at criticality. 
The study of dynamics of book sales \cite{sornette2004,deschatres2005dynamics} and YouTube videos views \cite{crane2008} has shown evidence of these critical triggering effects in large social networks. 
Open source projects and their online communication platforms coupled with the code repository serve a similar social network role yet at much smaller scales \cite{madey2002open,crowston2005social}.
Since these above analyses as well as those presented here benefit from the survival bias,
in other words the analyses are performed on top performers among a much larger database, 
the existence of criticality in these system can be interpreted as the signature of a degree of success
quantified by significant activity. Specifically, considering a large universe of projects, those that
are of interest in the sense of exhibiting significant dynamics in volume and quality are those
for which the conditions are met to be close to criticality.

\subsection{Large deviation mechanism  for superlinear production \label{lardevarg}}

\begin{figure}[h]
\centerline{\epsfig{figure=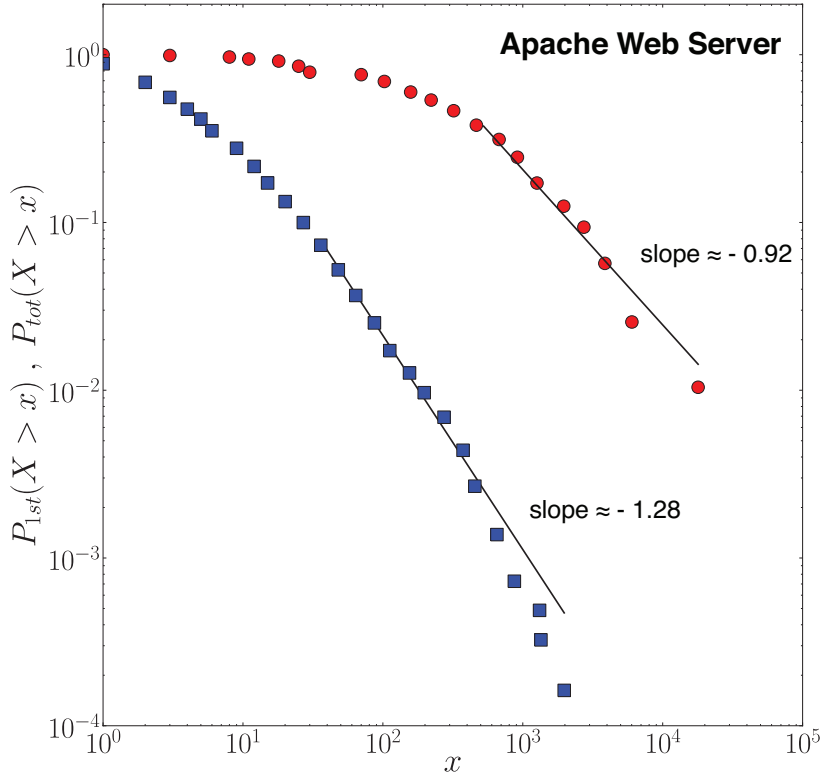,angle=0,width=12cm,scale=1}}
\caption{\footnotesize  {{\bf (blue squares)} Apache Web Server project: Complementary cumulative distribution $P_>^{1st}(r)$ of contributions (number of commits) per developer and per 5-day time bins (1st generation daughters events in the language of the epidemic branching process described in the text) with exponent $\gamma \approx 1.28$. {\bf (red circles)} Complementary cumulative distribution $P_>^{tot}(r)$ of all contributions per developer over a long period of time. $P_{tot}$ is equivalent to measuring the cluster sizes of contributions following critical cascades (\ref{sdfjhsg9e}). All distributions have been fitted using the maximum likelihood estimator (MLE). The distribution of cascade size is characterised by the exponent $\approx 0.92 <1$ compared to the first generation daughter events distribution with exponent $\gamma \approx 1.28$. The results showed here for Apache are representative of the distributions found in other collaborative projects.}}
\label{pwlaws}
\end{figure}

The second class of mechanisms builds on the evidence of 
large deviations in the statistics of the production activity $R$ over the whole
population of contributors and over the whole life of the project.
Figure \ref{pwlaws} shows the complementary cumulative distribution $P_>^{tot}(r) := {\rm Pr}(R> r)$ of all contributions per developer over a long period for the Apache Web Server project. One can observe
an approximate power law tail dependence 
\begin{equation}
P_>^{tot}(r) \sim 1/r^{\mu}~, 
\label{srthyjueyt}
\end{equation}
with $\mu \approx 0.92$. Within the epidemic framework presented in the next section, 
$P_>^{tot}(r)$ will be shown to be equivalent to the statistics of the cluster sizes of contributions 
following critical cascades \cite{saichev2005power} [see expression (\ref{thujyu})]. 
This result, showed for the Apache Web Server project, is representative of the distributions found in other collaborative projects.

In the presence of such a power law statistics of contributions characterized by an exponent $\mu <1$,
we show below that the sum of contributions over all developers is controlled by extreme contributors.
The contributions made by these exceptional members of the group are also responsible for the observed superlinear behavior given by (\ref{superlineqex}). This mechanism is reminiscent of the improved group performance that results
from the presence of one or few surperforming individuals \cite{shaw1932comparison}. In this case,
the largest contributor provides a finite fraction of the whole production over a given time period. 
This largest contributor (i.e. the ``large deviation'') has a superlinear contribution in the group size \cite{bouchaud1990anomalous,sornette2006critical}. In this situation, the increasing productive activity results 
from a large heterogeneity of activity per individual. 
And the more contributors $c$ during a production period, the more likely
it is to find an extremely large contribution. 

Specifically, starting from expression (\ref{srthyjueyt}) for the 
complementary cumulative distribution $P_>^{tot}(r)$, we denote 
$p(r) \sim 1/r^{1+\mu}$ the corresponding
probability density function obtained as the derivative of $P_>^{tot}(r)$.
Let us call $\{R_1, R_2, ..., R_{c-1}, R_c\}$, the total number of commits
contributed respectively by the developers $1, 2, ..., c-1, c$.
Let us call $R_{\rm max}(c)$, the largest among the set  $\{R_1, R_2, ..., R_{c-1}, R_c\}$.
A good estimate of $R_{\rm max}(c)$ is obtained by the condition that the
probability $\int_{R_{\rm max}(c)}^{+\infty}  p(r) dr$ to find a developer with a total 
contribution equal to or larger than $R_{\rm max}(c)$ times the number $c$ of active developers
is equal to $1$, i.e., by the definition of $R_{\rm max}(c)$, there should be typically only one 
developer with such a number of commits. This yields
\be
 R_{\rm max}(c)  \sim c^{1/\mu} ~. 
 \label{sdfjhsg9e}
\ee

An estimate of the typical total number of commits $R_1 +  R_2 + ... + R_c$ contributed
by the $c$ developers can  then be obtained as  \cite{bouchaud1990anomalous,sornette2006critical}
\be
R_1 +  R_2 + ... + R_c   \approx c  \int_0^{R_{\rm max}(c)} r   p(r) dr
 \sim  c^{1/\mu}    ~,~~~~{\rm  for}~\mu <1~.
 \label{rjtik6ik}
\ee
We stress that the scaling $\sim c^{1/\mu}$ only holds for $\mu <1$ and is replaced
by $\sim c$, i.e., linearity, for $\mu > 1$.
The upper bound in the integral in (\ref{rjtik6ik}) reflects that
the random variables $\{R_1, R_2, ..., R_{c-1}, R_c\}$ are not larger than $R_{\rm max}(c)$
by definition of the later. According to equation (\ref{rjtik6ik}), the typical
total production (number of commits) by $c$ developers
is proportional to $c^{1/\mu}$, when their contributions are wildly distributed
with a power law distribution with exponent $\mu <1$. According to this large
deviation mechanism, the superlinear exponent $\beta$ is equal to $1/\mu$.
\be
{\rm \bf prediction~of ~the~ large~ deviation ~mechanism}: ~\beta = 1/\mu~, ~{\rm for}~\mu < 1~.
\label{eyyn}
\ee

Within this large deviation mechanism, explaining the superlinear productive activity ($\beta>1$) 
reduces to explaining the heavy-tailed distribution of commits $R$ per contributor over a large period of time, i.e., amounts to derive the power law distribution (\ref{srthyjueyt}) with $\mu <1$. For this, the next section proposes a generic model.

\subsection{Cascading model of productive activity
\label{trhtyjitukiu|}}
Both the {\it interaction-based} and the {\it large deviations} mechanisms 
can be captured together by a generic cascade process, which is well described by the excited 
Hawkes conditional Poisson process \cite{hawkes1974acluster}.
The Hawkes process typically captures well a variety of social dynamics involving complex human interactions such as online viral meme propagation \cite{crane2008}, gangs and crime in large American cities \cite{mohler2011}, cyber crime \cite{baldwin2012} and financial contagion \cite{ait-sahalia2010,filiminov2012,filiminov2014}.  
The Hawkes process is defined by the intensity $I(t)$ of events (commits) given by
\be
I(t)= \lambda(t) + \sum_{i | t_t<t}  f_i \phi(t-t_i)~,
\label{jruym}
\ee
where $\{t_i, i=1, 2, ...\}$ are the timestamps of past commits, $\lambda(t)$ is the spontaneous
exogenous rate of commits, $f_{i}$ is the 
fertility of commit $i$ that quantifies the number of commits (of first generation)
that it can potentially trigger directly, and $ \phi(t-t_i)$ is the memory kernel, whose
integral is normalized to $1$, which weights how 
much past commit activities influence future ones. $\phi$ typically reflects how tasks are prioritized and performed by individuals according to a rational economy where time is a non storable resource \cite{maillart2011}. 
Expression (\ref{jruym}) expresses that the number of commits contributed between time $t$ and $t+dt$
results from two sources: (i) an exogenous source $\lambda(t) dt$ representing the spontaneous commits not related
to previous commits; (ii) an endogenous term represented by the sum over all commits that were 
made prior to $t$, and which are susceptible to trigger future commits. An obvious triggering mechanism
is debugging: a past commit may attract the attention of a developer who fixes a bug and thus improves the code.
Another triggering mechanism by which a previous commit may trigger a future commit is when the former
enables new functionalities and relationships that open novel options for the developers. 
The Hawkes model is the simplest conditional Poisson process that combines both exogeneity and
endogeneity.

The class of Hawkes models can be mapped onto the general class of branching processes \cite{daley2007}. 
The statistical average fertility $\langle f_i \rangle$ defines the branching ratio $n$, which is the key
parameter. For $n<1$, $n=1$ and $n>1$, the process is respectively sub-critical, critical and super-critical \cite{helmstetter2002subcritical,helmstetter2003}.
In the sub-critical regime ($n<1$), the average activity tends to die out exponentially fast and 
the exogenous source term $\lambda(t)$ controls the overall dynamics. At criticality ($n=1$), 
on average one commit is triggered in direct lineage by a previous commit, corresponding
to a marginal sustainability of the process with infinitesimal exogenous inputs.
The super-critical regime ($n>1$) is characterised by an explosive activity that can 
occur with finite probability. Interpreting a cluster or connected cascade in a given branching process of triggered contributions as the burst of production in a group of developers, 
the distribution of contributions is thus mapped onto that of triggered cluster sizes \cite{saichev2005power}.

Let us define
the complementary cumulative distribution $P_>^{1st}(r)$ of contributions (number of commits) per developer 
directly triggered by a given past commit, which can be called first-generation {\it daughter} commits generated by
a {\it mother} commit. Consider the case where $P_>^{1st}(r)$  is also a power law 
\be
P_>^{1st}(r) \sim {1 \over r^\gamma}~.
\ee
Close to or at criticality, the distribution of cluster sizes, which is equivalent to the distribution of productive activity $P(r > R)$ given by (\ref{srthyjueyt}) has an exponent $\mu=1/2$ \cite{harris2002theory}, under the condition that the distribution $P_>^{1st}(r)$ of contribution sizes triggered directly by previous contributions (so-called first-generation cascades) decays sufficiently fast, i.e., with $\gamma \geq 2$. The result  $\mu=1/2$ holds also for any distribution $P_>^{1st}(r)$ 
 decaying asymptotically faster than a power law \cite{saichev2005power}.
When $1<\gamma<2$, the  mean field exponent $\mu=1/2$ is changed into \cite{saichev2005power}
\be
\mu=1/\gamma~.
\label{thujyu}
\ee
Together with (\ref{eyyn}), the superlinear exponent $\beta$ is predicted to be
\be
\beta = 1/\mu =\gamma, ~~~~~~~~~~{\rm for}~ 1 \leq \gamma \leq 2~,
\label{eyyqethyujn}
\ee
that is, equal the exponent $\gamma$ of the tail distribution of first generation contributions 
per developers. For $\gamma >2$, $\mu=2$ and therefore $\beta=2$.
An analytical derivation of the prediction (\ref{eyyqethyujn}) 
using the Hawkes process (\ref{jruym}) that anchors rigorously the large deviation argument 
of Section \ref{lardevarg} is given by Saichev and Sornette \cite{saichev2014superlinear}.

\begin{figure}[h]
\centerline{\epsfig{figure=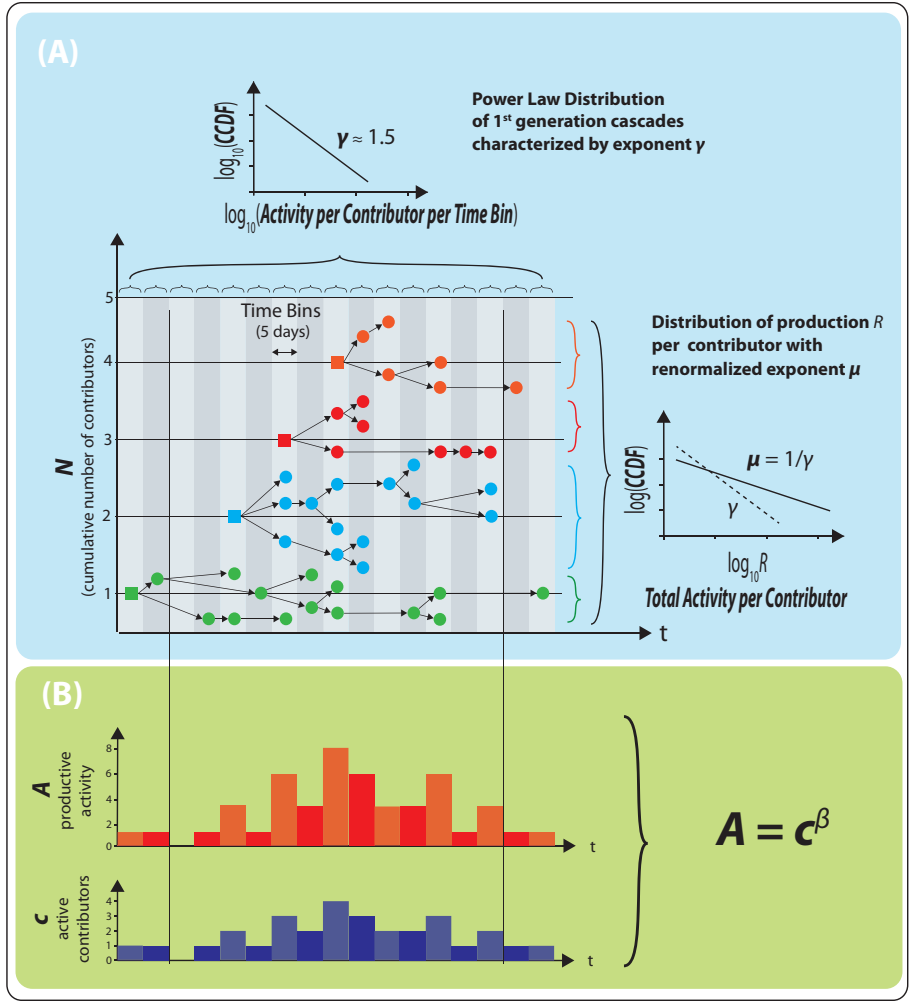,angle=0,width=12cm,scale=1}}
\caption{\footnotesize  {\bf (A)} (light blue) Triggering mechanism generating the clusters of size with renormalized exponent $\mu = 1/\gamma$ from the distribution of first generation ``daughter events" with exponent $\gamma$. For the sake of simplicity, we represented one cluster of activity per contributor, but triggering can occur between contributors provided that the probability of triggering remains the same between all contributors. {\bf (B)} (light green) shows how the triggering mechanism generates superlinear productive activity $A$ as a function of the number of active contributors $c$.}
\label{schema}
\end{figure}

Figure \ref{schema} synthesizes the relation between superlinear productive activity, (critical) cascades, the distribution of first-generation triggering and the total distribution of activity per contributors over a sufficient long period.

\subsection{Empirical tests}
We now turn to empirical tests of this theory.
For each $250$ days period, we have calibrated the power law tails of two distributions:
\begin{enumerate}
\item the distribution of the total number $r$ of commits per contributor over the $250$ days,
which is taken as a proxy for $P_>^{tot}(r)$, with exponent $\mu$;
\item the distribution of the number of commits 
per developer per $5$ days time bin, which is assumed to be a reasonable proxy for
the distribution $P_>^{1st}(r)$ of the first generation production characterized by 
the exponent $\gamma$.
\end{enumerate}
For each OSS project, we have used the discrete maximum likelihood estimator 
(MLE) with a p-value threshold $p > 0.1$, obtained by bootstrapping, 
and Kolmogorov-Smirnov Distance $KS < 0.15$ to select the ranges over which the
calibration is performed \cite{Clauset2009Powerlaw}. 

\begin{figure}[h]
\centerline{\epsfig{figure=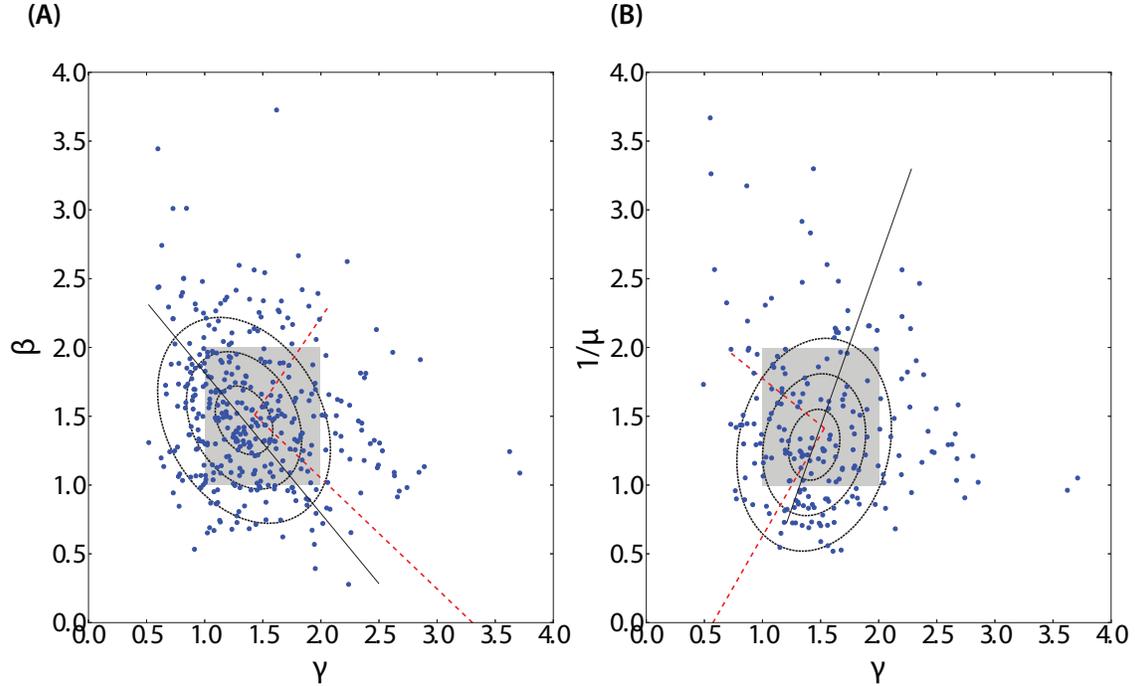,angle=0,width=15cm,scale=1}}
\caption{\footnotesize  {\bf (A)} superlinear exponent $\beta$ as a function of $\gamma$, the exponent of the power law tail distribution of first generation productivity for each of the $250$ days periods for which both values could be calibrated. The points are concentrated around $\beta \approx \gamma \approx 4/3$ with almost half of them ($184$ over $390$ values) within the grey area delimited by $1 \leq \beta \leq 2$ and $1 \leq \gamma \leq 2$. To test for the relations $\beta = \gamma$ and $1/\mu = \gamma$, we used a bi-Gaussian model. The dotted ellipses show the first three standard deviations around the barycenters and the black line represents the main axis with the bi-Gaussian model. We also performed a principal component analysis (PCA). The red dotted lines show the main direction of variance obtained with the PCA. Both methods show a positive relation between $\beta$ and $\gamma$ only on second principal component (slope $\approx 1.24$ with PCA). {\bf (B)} same as panel (A) for the dependence of $1/\mu$ versus $\gamma$ with a concentration of points in the grey area (86 over 213 values ) and $1/\mu \approx \gamma \approx 4/3$. Both the bi-Gaussian fit and the PCA show strong evidence of a positive relation with slope $\approx 1.02$ with
the bi-Gaussian approach and $\approx 1.47$ with the PCA}
 \label{bi-Gaussian}
 \end{figure}

Figure \ref{pwlaws} shows the result for the Apache Web Server project.
The fitting procedure qualifies the existence of a power law tail for the two empirical distributions
with estimated exponents respectively equal to $\mu= 0.92 \pm 0.1$ and $\gamma \approx 1.28 \pm 0.1$.
These values with their error bars are compatible with the prediction 
(\ref{thujyu}) $\mu =1/\gamma$, 
resulting from the cascades of triggering  \cite{saichev2005power}. This result is typical of the other
investigated OSS projects, as shown Figure \ref{bi-Gaussian}, albeit with a considerable variability.
This is expected since the projects are likely to be characterized by many more dimensions
that the production and cascading effects considered here.

Figure \ref{bi-Gaussian} presents $\beta$ as a function of $\gamma$ (panel A) and 
$1/\mu$ as a function of $\gamma$ (panel B) for all the OSS projects on our database,
According to the cascading model of productive activity of Section \ref{trhtyjitukiu|},
we should have $\beta = \gamma = 1/\mu$, according to (\ref{eyyqethyujn}).
Indeed, one can see that $\beta$, $\gamma$, and $1/\mu$ are clustered around $\approx 4/3$. 
Almost half of the considered periods ($184$ of a total of $390$) fitted over all projects 
belong to the regime where $1 < \beta < 2$ and $1 < \gamma < 2$ (panel A) 
and forty percent (86 out of 213) are such that  $1 < 1/\mu < 2$ (panel B) 
as predicted by the theory. 

Let us first focus on the relationship between $1/\mu$ and $\gamma$ shown in panel B of 
Figure \ref{bi-Gaussian}. Note that the statistics on the exponent $\mu$ is significantly smaller
compared to that for $\gamma$ simply because we obtain one data point over each $250$ day
periods for $\mu$ compared with one data point per $5$ days time bin for $\gamma$.
The shaded square represents the domain over which the theory applies (86 over 213 data points).
To test quantitatively the relation $1/\mu = \gamma$, we used a Gaussian bivariate
distribution model. 
The dotted ellipses show the first three standard deviations equi-levels around the barycenter
$1/\mu \approx \gamma \approx 4/3$ and the black line  represents  the principal axis of the bi-Gaussian model. 
We also performed a principal component analysis (PCA). The red dotted lines show the two
main directions of the variance obtained with the PCA. Both methods support a positive correlation between $\beta$ and $\gamma$ 
with slope $\approx 1.02$ with the bi-Gaussian approach and $\approx 1.47$ with PCA. 
To our knowledge, this may be the first empirical test ever of the renormalization of the exponent
$\gamma$ of first generation events into the renormalized exponent $\mu = 1/\gamma$
due to the cascade of triggering over all generation in a critical branching process
\cite{harris2002theory,saichev2005power}.

The evidence  for the relationship 
between $\beta$ and $\gamma$ is presented in panel A of Figure \ref{bi-Gaussian}. 
First, one can observe a prevalence of the large-deviation critical interaction regime as the 
grey square area delimited by $1 \leq \gamma, \beta \leq 2$ is very densely populated 
(184 out of 390).
Second, as already pointed out, the barycenter of the cloud of data points is on
$\beta \approx \gamma \approx 4/3$, as expected from theory.
However, we find limited support for a clear linear relation between $\beta$ and $\gamma$. 
The bi-Gaussian model analysis provides the three dotted ellipses 
showing the first three standard deviations away from the barycenter.
The black line representing the main axis of the bi-Gaussian model suggests
a negative correlation between $\beta$ and $\gamma$. Using a PCA analysis, we find
a positive relationship on the second principal component, with slope $\approx 1.24$.
These results suggest that very productive projects and periods within projects, 
characterized by a large superlinear exponent $\beta$, are likely to be due to more
complex interactions between the developers and their mutual triggering that
assumed by the simple theory developed above. In particular, differentiation 
between same-developer commit triggering and inter-developer commit triggering
seem necessary along the lines of Refs. \cite{saichev2011,saichev2013hierarchy}.
 
\section{Reconciling present findings and superlinear production in large cities}
\label{sdfuishfqsatug}

\begin{figure}[h]
\centerline{\epsfig{figure=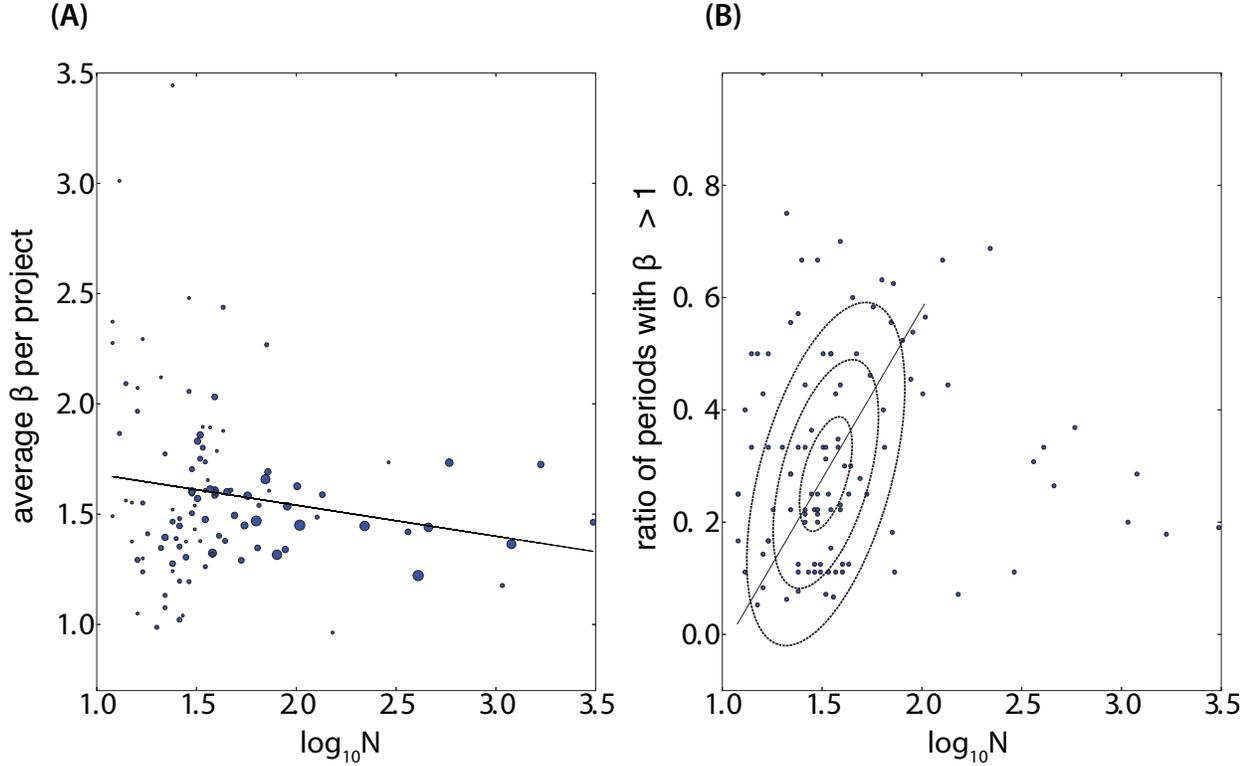,angle=0,width=17cm,scale=1}}
\caption{\footnotesize  {{\bf (A)} Average superlinear exponent $\langle \beta \rangle$ per project as a function of the cumulative number of contributors. The circle size reflects the number of exponents fitted per $250~days$ time window, for each project and entering the average statistics. The sampling ranges from $1$ (small disks) to $16$ (largest disk). $\langle \beta \rangle$ exhibits a slightly negative slope $\approx -0.14$ as a function of $\log_{10}(N)$ ($p<0.1$ and $r=-0.17$). {\bf (B)} To measure the prevalence of productive bursts in projects, we measure the ratio of periods with superlinear exponent $\beta >1$ over all $250~days$ periods for each project as a function of $\log_{10}(N)$}. We distinguish a cluster of points around  $\approx 0.3$ and $\log_{10}(N) \approx 1.52$ (i.e. $N \approx 33$ contributors) with a positive relationship ($slope \approx 1.37$) of the $Ratio$ as a function of $log_{10}(N)$. Projects with a large pool of contributors ($N >100$) are more randomly scattered with a lower ratio and do not obey the same relationship, suggesting a different regime.}
\label{agg_stats}
\end{figure}

Figure \ref{agg_stats} reveals that the clouds of superlinear production exponent $\beta$
exhibit an interesting regularity as a function of the total number of contributors $N$ of an OSS project.
The intuition motivating this investigation is the following. While a minimum critical mass 
of contributors is needed to foster productive bursts, large projects suffer from coordination costs, 
which may offset the increasing return of productive activity. 
Figure \ref{agg_stats} (panel A) shows indeed that the superlinear exponent $\beta$ decreases 
on average with the size of the projects . 
Panel B demonstrates that, for projects of up to $33$ contributors, the number  of $250$ 
days periods with $\beta > 1$ (superlinear regime) increases as a function of the total number $N$
of developers, approximately according to 
\be
({\rm ratio~of~time~windows~with}~\beta >1)    \sim  1.37 \log_{10}N~.
\ee
For $N>33$, a different regime occurs characterized by a much smaller ratio of the time periods
with superlinear productivity ($\beta > 1$).
Taken together, the two panels of Figure \ref{agg_stats} support the view that superlinear 
productivity is the appanage of relatively small projects with no more than 30-40 developers 
in total, while larger groups face the difficult challenge of creating and maintaining productive bursts.
The data is too scattered unfortunately to allow us to draw a firm conclusion
on the value(s) that $\beta$ converges towards for large project sizes.

There may be a link between our results and a previous study reporting
 the phenomenon of superlinearity on a completely different class of objects, namely cities.
Data from 360 US metropolitan areas have shown that 
wages, number of patents, GDP and intensity of crime scale superlinearly with population size
[\nobreak production $\sim ({\rm population})^\beta$ ] with an exponent $\beta \approx 1.15$ 
\cite{bettencourt2007,bettencourt2010}.
The value of $\beta$ larger than $1$ reflects the fact that productivity increases by about
11\% with each doubling in population \cite{bettencourt2010urban}. Qualitatively in line with our findings, 
the superlinearity found in our OSS data is significantly stronger ($\beta \approx 4/3$
on average, with 
large variations and some projects being characterised by much larger $\beta$'s) 
for the smaller projects with no more than 30-40 developers. We note that our results
apply to a completely different range of group sizes compared with the results for cities 
involving population of tens of thousand to tens of  millions inhabitants. 

The underlying mechanisms are perhaps different  \cite{bettencourt2013origins}. For cities, the superlinear scaling
in urban productivity demonstrates the importance of cities as centers of enhanced
interactions, leading to generation and exchange of knowledge and exploitation
of innovations \cite{bettencourt2010urban}. For the OSS projects, many other factors
come into play, such as the role of diversity and complementarity, which describes the fact that 
doing more of one thing increases the return to doing more of another.
Other possible mechanisms include synergies, economies of scale, 
coordination and leadership, role model and entrainment effect,
motivations,  friendship and other psychological factors.
However, Figure \ref{agg_stats} suggests that these mechanisms dampen out
as the project size becomes very large, possibly leaving only those still active
at the level of city sizes. 

Expanding on the remark on the different sizes involved in our OSS database compared
with cities, we present a simple mechanism and theoretical argument that may
explain the smaller value of the superlinear exponent for cities, deriving it
from our results obtained for small group sizes.  The key idea is that the population of 
a city can be partitioned into many groups of persons interacting closely within a group
and loosely or not at all across groups. Groups can be firms, or department within firms,
clubs, and other organisations through which people interact.
We assume that, within each group, the superlinear production law 
(\ref{superlineqex}) holds with the exponent $\beta \approx 4/3$ found in our OSS database.
 
 The second ingredient is that group sizes $g$ are widely distributed, roughly as Zipf's law \cite{saichev2009},
 \be
 p(g) \sim {1 \over g^{1+z}}~,
 \label{trhrujkiu}
 \ee
 where $p(g)$ is the probability density function of the group sizes $g$,
 $z=1$ if Zipf's law holds exactly, while in general $z$ can deviate from $1$ for
 a variety of reasons \cite{malevergne2013zipf}. 
 Let us assume that a city of total population $N$ is constituted of $n$ groups, respectively with memberships of 
$N_1, N_2, ..., N_n$ individuals. The total production of the city is then, according to (\ref{superlineqex}),
\be
R(N) = N_1^\beta + N_2^\beta + ... + N_n^\beta   ~,
\label{rjryumki}
\ee
assuming for the moment and for simplicity that $\beta$ is independent of group sizes.
$R(N)$ in expression (\ref{rjryumki}) can be estimated as \cite{bouchaud1990anomalous,sornette2006critical}
\be
R(N) \sim   n  \int_1^{g_{\rm max}(n)}  g^\beta ~p(g) dg~,
\label{jryukieyw}
\ee
where $g_{\rm max}(n)$ is the largest group size among the $n$ groups, which can be estimated by
\be
n \int_{g_{\rm max}(n)}^\infty p(g) dg \sim 1  ~ ~~ \to ~~  g_{\rm max}(n) \sim n^z~.
\ee
By conservation and assuming for simplicity no strong overlap between the groups,
we have approximately
\be
N_1   + N_2 + … + N_n  =  N  \sim  n  \int_1^{g_{\rm max}(n)}  g ~p(g) dg~.
\ee
This leads to $n \sim N$ for $z>1$ and $n \sim N^z$ for $z<1$. In words, 
a relatively thin tail of the group size distribution ($z>1$) is associated with a number
of group scaling proportionally to the total city population $N$. In contrast, for a heavy
tailed distribution ($z<1$), the number of groups scales sublinearly with $N$, as
the few largest groups account for a finite fraction of total population. Reporting in expression (\ref{jryukieyw}),  this yields $R(N) \sim N^b$, with the exponent $b$
obeying three possible regimes.
\begin{enumerate}
\item $z \leq 1$ implies $b = \beta$:   the same superlinear production exponent defines
the whole city production as a function of its population as does the production of each 
independent group. The mechanism is clear: for $z<1$, a few single largest
groups dominate the $n$-partition and account for the majority of the city population.
The same scaling holds essentially because the city is almost controlled by a single group
and we have assumed the same exponent $\beta$ for all groups. The empirical
evidence suggests that this case does not apply.

\item $1 < z < \beta$ implies $b = \beta/z$.  In this regime, there are still very large groups that
contribute to the superlinearity but their relative numbers is much less than for $z \leq 1$.
The values $\beta=4/3$ with $b = 1.15$ 
can be reconciled with $z= \beta / b \approx 1.16$. This exponent is, with error bounds, 
roughly compatible with the value found for firms in the US, close to $1.25$ \cite{ramsden2000companysize}.
 
\item $1 < \beta < z$ implies  $b=1$, which corresponds to a linear growth of production of the
city with its population. In this regime, the overall city production is controlled by
the many small groups constituting the city and there are no scale effects other than
a proportionality with the number of small groups.
\end{enumerate}

While this argument is quite naive, it demonstrates the importance of 
the interplay between partitions of cities in groups, the corresponding productivity
of such groups and the size distribution of these groups.
A similar story is likely to be relevant in large OSS projects, groups and firms, which 
for a variety of reasons ranging from cognitive limitations \cite{zhou2005dho} 
to efficiency maximization \cite{toulouse1978}
are found to organize in subgroups, often in a hierarchical way \cite{zhou2005dho}.

\section{Discussion}
\label{sdfudprtptm}
In the early days of the industrial revolution, Adam Smith noted how the successive efficiency gains of communication means have helped reach unprecedented pools of resources and how they have unlocked some limitations of the labor market through improved division of labor \cite{smith1776}. The telegraph, telephone and more recently the
Internet have further pushed back the possibilities for knowledge production and for labor organizations on the model of collective action \cite{ostrom1990}. Nowadays, unrelated people spontaneously team up across the world in open collaboration projects and join forces to create knowledge in the form of software, natural language \cite{wuchty2007}, mathematics \cite{gowers2009} as well as for the production of tangible goods \cite{raasch2009}. These organizations rely primarily on the principles of peer-production \cite{benkler2002}: (i) task self-selection, (ii) peer-review and (iii) iterative improvement, at odds with traditional market and firm production organizations \cite{coase1937}. Expertise can be timely and rightly pulled from a broader community towards efficient problem resolution. The present understanding of group performance in social psychology goes in the same direction: experiments involving small groups performing coordination tasks \cite{ingham1974,tziner1985}, problem solving \cite{shaw1932comparison} and innovation 
\cite{sacramento2006team} support the hypothesis that larger groups perform better because more diverse cognitive abilities can be pooled. Group productive activity can also be  {\it more than the sum of their parts} if members develop social sensitivity  among each others \cite{woolley2010}. However, the marginal gain of having more individuals in a group decreases rapidly to be negligible beyond five individuals \cite{gordon1924group,shaw1932comparison,laughlin1966}. 
Similarly, as projects attract larger communities, more coordination is required through social norms and formal governance structures\cite{O'Mahony2007}, which may in turn reduce the positive effects of peer-production \cite{halfaker2013}.

\section{Conclusion}
\label{sdfiuetgnzgposfpkn}
In this paper, we have shown that productive bursts, associated with increasing return of activity, result from the mechanism of critical triggering of commits among contributors. Such critical triggering
may operate according two co-existing mechanisms: {\it interactions} and {\it large deviations}.  These mechanisms 
have been falsified in three independent ways : (i) documenting the superlinear relationship between productive activity $R$ and the number of active contributors $c$ characterized by the scaling exponent $1 < \beta < 2$; (ii) measuring
the power law tail distribution of first generation cascades  with exponent $1 < \gamma < 2$ and checking
that it explains the superlinear productivity exponent $\beta$; 
and (iii) measuring the power law tail distribution of production cluster sizes with exponent $\mu$ and verifying that it is 
approximately equal to the $1/\gamma$, where $\gamma$ is the distribution of contributions per developer
at short times.

We have found that superlinear productive activity holds for a broad range of project sizes and types, with a slight decrease of the average scaling exponent $\beta$ with the total number of contributors $N$. The frequency of productive bursts occurrence in projects has
been found to be very large for $N \leq 33$ compared with larger projects. The results suggest that size and threshold effects have an influence on the ability to trigger and maintain critical triggering of individual contributions. Indeed, 
contributions must create enough reaction opportunities to trigger on average as many 
downward contributions. Pervasive communication systems (social networks), physical proximity (e.g. cities), or even personal dedication to the project surely help increase opportunities for a contribution 
to trigger a follow up action. On the other hand, large and complex structures with overwhelming communication
loads or inadequate governance structure can inhibit the {\it ripe} circulation and reuse of knowledge for the sake of further cumulative innovation. The {\it large deviation} mechanism provides another take-away lesson: open collaboration does not imply equal work between contributors. On the contrary, productive bursts are the hallmark of 
a minority of individual engagement with intense interactions and short-lived contributions
of far above average sizes. Whether these large deviation contributions pull engagement by others or on the contrary are pushed by the community remains an open question to be elucidated.

\vskip 0.5cm
\noindent
\acknowledgements{One of the authors (T.M.) acknowledges support from the Swiss National Science Foundation (Grant Nr.  PA00P2-145368).}

\pagebreak

\vspace{1cm}

\clearpage

\vspace{1cm}

\clearpage


\clearpage

\vspace{1cm}

\clearpage

\vspace{1cm}

\clearpage

\vspace{1cm}

\clearpage

\vspace{1cm}

\clearpage

\vspace{1cm}

\end{document}